# Inversion of 1D Frequency- and Time-domain Electromagnetic Data with Convolutional Neural Networks

Vladimir Puzyrev[1,*] and Andrei Swidinsky[2]

[1] *School of Earth and Planetary Sciences and Curtin University Oil and Gas Innovation Centre, Curtin University, Perth, WA 6102, Australia.*

[2] *Department of Geophysics, Colorado School of Mines, Golden, CO 80401, USA.*

**Abstract**

Inversion of electromagnetic data finds applications in many areas of geophysics. The inverse problem is commonly solved with either deterministic optimization methods (such as the nonlinear conjugate gradient or Gauss-Newton) which are prone to getting trapped in a local minimum, or probabilistic methods which are very computationally demanding. A recently emerging alternative is to employ deep neural networks for predicting subsurface model properties from measured data. This approach is entirely data-driven, does not employ traditional gradient-based techniques and provides a guess to the model instantaneously. In this study, we apply deep convolutional neural networks for 1D inversion of marine frequency-domain controlled-source electromagnetic (CSEM) data as well as onshore time-domain electromagnetic (TEM) data. Our approach yields accurate results both on synthetic and real data and provides them instantaneously. Using several networks and combining their outputs from various training epochs can also provide insights into the uncertainty distribution, which is found to be higher in the regions where resistivity anomalies are present. The proposed method opens up possibilities to estimate the subsurface resistivity distribution in exploration scenarios in real time.

[*] *Corresponding author, vladimir.puzyrev@gmail.com*

## 1. Introduction

The meteoric rise of deep learning (DL) models in many fields of science and technology has attracted considerable attention from the geophysical community and made artificial intelligence one of the main focuses of attention from both academia and industry. Various types of deep neural networks have shown great promise in applications such as the detection of faults (Araya-Polo et al., 2017; Huang et al., 2017; Wu et al., 2019), seismic facies classification (Zhao, 2018; Souza et al., 2019), digital rock physics (Mosser et al., 2018; Sudakov et al., 2019; Karimpouli and Tahmasebi, 2019) and many other geological problems. The main advantages of DL methods are that they allow the detection and exploitation of nonlinear dependencies in the data without specifying a particular model in advance. In order to achieve maximum performance, DL methods require massive amounts of data, which makes them well-suited for handling large-scale datasets. They also scale better with problem size compared to other machine learning techniques, and in many cases are able to produce better results. Features automatically learned by DL may potentially be more valuable and better suitable for analysis than hand-engineered features. By exploiting different layers of abstraction, deep neural networks are able to discover both the low-level and high-level features in data.

Inversion of electromagnetic (EM) data is an important problem with applications in different areas of geophysics. Numerical methods for 1D EM inversion have been extensively studied in the past decades, e.g., Farquharson et al. (2003), Auken et al. (2005), Constable and Weiss (2006), Key (2009), Moghadas et al. (2015), Pardo and Torres-Verdín (2015). Nowadays 1D methods find a place in EM interpretation where 2D or 3D inversion is not possible due to lack of data or high computational cost. Modelling in horizontally layered media often employs semi-analytic methods that are advantageous in terms of accuracy and computational efficiency (Løseth and Ursin, 2007; Streich and Becken, 2011; Swidinsky et al., 2018). Maxwell's equations in this case permit a decomposition of the EM field in two independent modes, namely a transverse electric (TE) mode and a transverse magnetic (TM) mode, and the resulting field is simply the sum of these two modes (Ward and Hohmann, 1988).

The inverse problem in non-unique, i.e. many models fit the data equally well. Local optimization methods for inversion such as the nonlinear conjugate gradient or Gauss-Newton are highly prone to getting trapped in a local minimum, especially when the starting model is chosen far away from the optimal model. Inversion of large-scale geophysical data often requires the use of high performance computing systems, especially when measurements at multiple source-receiver positions are considered (Puzyrev et al., 2018). Probabilistic inversion methods have been applied to various EM problems as well (Gunning et al., 2010; Minsley, 2011; Ray et al., 2013; Wang et al., 2019) though being computationally demanding even in low-dimensional setting.

Neural networks have been applied to the problems of estimation of the properties or spatial location of a simple target from geophysical data since the 1990s (Poulton et al., 1992; Röth and Tarantola, 1994). However, the level of the technology was very limited at that time and the method was only marginally applied in later years. Recent applications of deep neural networks to inverse problems include seismic (Araya-Polo et al., 2018; Richardson, 2018; Liu and Grana, 2019), electromagnetic (Puzyrev, 2019; Oh et al., 2019; Li et al., 2019), and X-ray computed tomography (Ye et al., 2018). These methods are mainly based on convolutional neural networks (CNN) which have shown tremendous success in image classification problems and are now universally used in image and video processing. One of the main advantages of the inversion based on DL is that it delivers the results in real time by predicting the distribution of subsurface properties from input data in a single step.

In this study, we explore the potential of deep CNN for 1D EM inversion. This approach to inversion is entirely data-driven, does not employ traditional gradient-based techniques and provides a guess to the model instantaneously. Our research provides two contributions to the field: (a) development of 1D CNN-based models for EM inversion and (b) application of these models to both synthetic and real datasets. In particular, we demonstrate the CNN approach on two types of electromagnetic survey measurements: marine frequency-domain controlled-source electromagnetics (CSEM) and time-domain electromagnetics (TEM). The former method is typically used for offshore hydrocarbon exploration while the latter is more commonly employed in mineral exploration, environmental engineering problems and hydrogeological studies.

This paper is organized as follows. First, we formulate the EM problems of interest in Section 2. In Section 3, we describe the architecture of the deep neural networks employed and the training process. The performance of the method is investigated on two examples including synthetic and real EM data in Section 4. Finally, the last section summarizes the outcomes and points out future research directions.

**2. Background**

The marine frequency-domain CSEM method measures the electric and magnetic fields on the seafloor which are produced by a grounded electric dipole antenna carrying alternating current. This type of transmitter generates vertical current flow in the seafloor and the method is therefore used to detect and image resistive targets like hydrocarbon reservoirs. In a typical survey, multiple EM receivers measuring up to three electric and three magnetic field components are deployed on the seafloor and the electric dipole antenna – transmitting multiple frequencies – is towed over the array; measurements are therefore made at transmitter-receiver offsets ranging from 100s of meters to 10s of kilometers. While 3D anisotropic inverse modelling is now the standard interpretation tool in such multi-offset, multi-component, multi-frequency CSEM surveys, we demonstrate our CNN methodology using synthetic data generated from 1D isotropic layered models; extension to 3D will be the subject of a future paper. A large amount of training data is one of the key requirements for DL methods. In order to accurately establish a relationship between the data and model parameters, the networks are required to see as many as possible different examples during training. To create such large database of synthetic CSEM responses, we use a parallelized version of the algorithm described in Swidinsky et al. (2018) to calculate the EM response of a CSEM system above a multi-layered seafloor.

In the TEM method, steady current in an ungrounded transmitting loop of wire is abruptly switched off, inducing secondary currents to flow in the subsurface by Faraday's Law. These secondary currents are primarily sensitive to conductive targets (such as massive sulfide mineralization and freshwater aquifers) and through Ampere's Law produce a secondary magnetic field which decays with time. The time-derivative of this magnetic field is typically measured as a decaying voltage using a vertical induction coil receiver placed at the center of the transmitting loop. In this configuration, it is still very common to use 1D inversion methods to interpret survey data, because the measurements are generally sensitive to the subsurface structure directly beneath the transmitter. Layered models of conductivity versus depth from multiple TEM soundings are stitched together to produce a pseudo-2D or pseudo-3D image of the subsurface. Our results using neural networks trained on 1D models are therefore directly applicable to present-day field practice and we demonstrate the CNN methodology on a real ABEM WalkTEM dataset from Denmark, provided to us by Guideline Geo. We use a modified version of the algorithm described in Swidinsky et al. (2012) to calculate the EM response of a TEM system above a multi-layered earth.

## 3. Methodology

*3.1 Inversion with neural networks*

The goal of the DL inversion is to identify the subsurface model directly from the given data in a single step without constructing the gradients. The workflow consists of three stages: (1) generation of data, (2) training of the network and (3) prediction of model parameters. While the first stage can be challenging in 2D and 3D conditions, the generation of 1D data typically does not pose significant problems due to the low number of model parameters and the availability of fast numerical simulators. Once a sufficiently large and representative set of measurements and corresponding resistivity models is generated, it can be used in network training. After training, the network is expected to correctly predict the resistivity distribution both for the data used in training (training set) as well as for new, previously unseen data (test set).

CNN are arguably the most common type of deep neural networks and one of the most powerful developments in artificial intelligence in recent decades. They can be viewed as a natural extension of neural networks for processing images and data with a grid-like topology. CNN have an ability to learn filters that represent repeating patterns. These networks have shown good performance in data processing tasks where picking the most important features from segments of data regardless of the specific location of these features within the data is essential. Deep CNNs include multiple convolutional layers, thus forming a hierarchy of feature detection. As the features extracted at the previous level become the input at the next level, the model is able to capture small- and large-scale features. Besides their natural applications in computer vision, social media, robotics, and autonomous vehicles, CNN are actively extended to other types of problems such as modelling and simulation of physical systems or predictive analytics.

Figure 1 shows the general architecture of the neural networks used in this paper. The input data (either frequency- or time-domain EM responses) is first processed by several CNN-pooling cascades and the resulting features are given as input to dense (fully-connected) layers. This is perhaps the most common CNN architecture which is commonly applied in computer vision applications (e.g., Simonyan and Zisserman, 2014; He et al., 2016), especially for classification tasks.

As input, the network takes an M x C array, where M is the number of input points and C is the number of channels. Each channel is an observation of a different quantity at M points in space (i.e. measurements at the receivers in CSEM) or time (in TEM). Various channels represent the electric and magnetic field components, amplitude and phase, frequency, etc. Convolutional layers in the first part of the network process low and high-level features in data. There are four levels separated between each other by pooling layers, which make the representation approximately invariant to small translations of the input and shrink the representation size, thus also reducing the computational cost. Each level has two or three convolutional blocks with convolution, batch normalization, and activation. Batch normalization (Ioffe and Szegedy, 2015) is used between the convolutional layers for improving the training performance and regularization purposes following recent trends in deep CNN architectures (He et al., 2016). Leaky rectified linear units (ReLU) are chosen as activation functions and a dropout of 0.1 (Srivastava et al., 2014) is applied after the hidden layers in the dense part of the network.

The number of filters used in convolutions increases by a factor of 2 after each pooling layer to enable the network to better learn features at higher abstractions. The first level has N filters; the particular choice of N is made depending on the number of field components in the input data as described in Section 4. The second part of the network consists of fully-connected layers that are well-suited for regression tasks. The output layer has the same number of neurons as there are unknowns in the resistivity model. In this study, we consider an isotropic case with fixed layer boundaries, thus the number of unknowns is equal to the number of layers in the model. Unknowns are chosen as the logarithm of resistivity to decrease the range of their values. The size of the output can be chosen independently of the input size, which means no strict restrictions on the dimensions and size of the data and model parameters. For comparison purposes, we also train several fully-connected networks and evaluate their accuracy and generalization potential versus CNN.

The main choices regarding the architecture of the network are (a) the depth of the convolutional part of the network and number of filters at each level and (b) the number of dense layers and the number of neurons at each layer. Networks which are too deep typically achieve very good performance in training but generalize poorly due to overfitting to the training data, thus requiring the usage of some regularization techniques. The present approach has several differences from the networks employed in Puzyrev (2019) for 2D EM examples. The latter method used the fully convolutional architecture to output a high-dimensional structured object and was adapted for monitoring scenarios (for example, taking as input the difference with the baseline subsurface model). While the method allowed for 2D pixel-based inversion with 10,000 unknowns, the monitoring setup is such that all the 20,000 sample models has one (or several located close to each other) resistive targets. The exploration setup is more challenging as there is limited information about the resistivity distribution and thus many more situations should be considered.

During training, the parameters (weights) of the network are progressively updated by minimizing the error on the training dataset (training error). We are interested though in how well the method generalizes to the data it has not seen before (test error) since this determines its practical performance. In order to choose optimal values for the hyperparameters of the network and to stop training before the network starts overfitting to the training data, we monitor the error on the validation dataset (validation error) during the training and choose the network that achieves the lowest validation error. This early stopping criterion allows avoiding overfitting to the training data.

Nesterov-accelerated adaptive moment estimation (Nadam) algorithm (Dozat, 2016) is used for error minimization and the root mean squared error (RMSE) over the training dataset is used as the loss function. Open-source libraries TensorFlow (Abadi et al. 2016) and Keras (Chollet et al. 2015) are used for implementation.

*3.2 Data generation*

One of the key requirements of DL methods is a large amount of data required for training. In order to accurately establish a relationship between the data and model parameters, we train our networks on multiple synthetic examples. For the frequency-domain marine CSEM inversion, we create a synthetic dataset consisting of 512,000 examples. The generation of this data took 25 hours on 8 computational nodes equipped with 28 Intel Xeon E5-2680v4 cores each using MPI-parallelized modelling algorithm of Swidinsky et al. (2018). The validation and test sets consist of 5120 examples each and the remaining 501,760 examples are used in training, which corresponds to a 98/1/1 split. This relatively large number of different examples used in training is expected to help us to achieve high generalization and low test set errors without data augmentation or adding strong regularization to the network.

The TEM inversion is less data-intensive compared to the marine CSEM method and the number of unknown parameters (i.e. the number of layers in the model) we consider in the TEM example below is 50 (against 100 in the CSEM example). We thus use a smaller dataset and study how the choice of regularization and network hyperparameters may improve generalization abilities. The TEM dataset consists of 10,240 examples; 9216 of them are used for network training and the remaining 1024 are split equally between the validation and test sets, which corresponds to a 90/5/5 split.

Both the CSEM and TEM input data are standardized using the mean $\mu$ and standard deviation $\sigma$ values on the corresponding training sets:

$$\tilde{x} = \frac{x - \mu}{\sigma},$$

and natural logarithm of the resistivity $m = \ln(\rho)$ is used as the unknown model parameter.

## 4. Numerical results

### 4.1 Frequency-domain CSEM inversion

In the first model scenario, we consider a marine CSEM setup where low-frequency EM energy is generated by a towed electric dipole. Following the well-studied model described by Key (2009), we use either two frequencies of 0.1 and 1 Hz or five frequencies of 0.1, 0.3, 1, 3, and 10 Hz. An inline (x-oriented) horizontal electric dipole generates $E_x$, $B_y$, and $E_z$ field components; only the horizontal electric and magnetic components are used in the inversion. The transmitter is located 25 m above the seafloor and the responses are computed at 50 m intervals at distances up to 20 km from the receiver. Thus, the input dimensions are $400 \times 4N_f$, where $N_f$ is the number of frequencies.

The resistivity models are inspired by real marine CSEM exploration scenarios. Each model has 100 thin layers starting from the seafloor; their thicknesses vary from 20 m near the seafloor to 100 m at depths of 4 km. Resistivity is assumed to be isotropic though an extension of the method to VTI-anisotropy is straightforward. Resistivity increases with depth from approximately 0.5–2 Ohm-m to 20–50 Ohm-m and has small random variations in each layer. Besides that, up to four resistive and one conductive anomalies are present in each model. Their details are summarized in Table 1.

Figure 2 compares the CNN predictions with the true models from the test dataset using all the five frequencies. The final normalized RMSE for the 5120 examples of the test dataset is 0.093. We also report the coefficient of determination ($R^2$), which takes values close to 1 when the estimated values closely correlate with the actual ones; in this example, the best $R^2$ value achieved on the test dataset is 0.912. In most cases, for models with several resistive layers the neural network detects the presence of all of them, although deeper targets are sometimes hidden by shallow resistive layers (examples (c) and (g) where the 3-km-deep targets are not detected). The thin conductive target located at a depth of 2.5 km is also shadowed by the shallow resistive anomaly in example (f). 35.6 per cent of the predictions have normalized RMSE lower than 0.08 and only 3.5 per cent have RMSE higher than 0.15. Omitting resistive or conductive anomalies results in poor RMSE and $R^2$ metrics as shown in example (i) for a model with gradually increasing resistivity and a 10 Ohm-m half-space.

In Figure 3, we show the inversion results using only two frequencies of 0.1 and 1 Hz. Using less data and lack of high frequency content impacts our ability to determine accurately the resistivity in the cases where several thin targets are clustered together, such as in example (h). The RMSE and $R^2$ metrics for most of the cases are inferior compared to the 5-frequency scenario, although the main features of the models are still determined correctly.

For the networks used in these examples, N is chosen as 80 and 32 for the 5- and 2-frequency cases, respectively (4 filters in the first convolutional layer per each input channel). The last layer thus has 800 filters for $N_f = 5$. The fully-connected part of the network has 6 hidden layers with 256 neurons each. The output layer has 100 neurons, each of which yields the natural logarithm of resistivity in the corresponding layer. The network has 8.5 millions of trainable parameters in total (4.1 million in the convolutional layers and the rest in the fully-connected layers). 0.093 is the best normalized RMSE achieved among different networks employed in this study.

To get a better understanding of the potential accuracy of the CNN-based inversion, we investigate the performance of the neural networks trained on datasets of different sizes ranging from 10 thousand to half a million examples. Table 2 shows the training and test errors for the CNN used above and a fully-connected network with 7 hidden layers each having 512 neurons. Using small training datasets results in a large gap between the training and test error (the networks overfit the data but generalize poorly). Several hundreds of thousands of training examples seem to be an optimal number needed for accurate and reliable prediction of the resistivity parameters in this scenario. The convolutional architecture achieves lower test set errors compared to the fully-connected network.

CNN inversion can also provide some insights into uncertainty distribution. Figure 4 compares the inversion results when using two different CNNs at 10 consecutive epochs (chosen from the range when the validation errors reach their plateau and overfitting to the training data might have started). CNN 1the same as used in the previous examples and CNN 2 has half as many filters and smaller kernel size. Three of the eight models shown previously, namely, (a), (e) and (f), show small variations in the resistivity estimation and thus are omitted from Figure 4. For the five remaining cases, the largest resistivity variations are observed near the anomalous regions. The deeper network (CNN 1) has superior performance and does not produce fluctuations at depths of 300–700 m as the shallower network (CNN 2) does. This suggests that using several deep CNN with slightly different hyperparameters and comparing their results from several training epochs can be used as a rather simple way to quantify uncertainty in the inverted models.

Figure 5 shows the inversion results for a 2D model comprised of 100 1D resistivity profiles. The data is obtained by 100 independent 1D simulations. These 1D models are parametrized in a slightly different way than the one described in Table 1 (lower baseline resistivity and smaller random fluctuations); nevertheless, the inversion results are accurate and the RMSE and $R^2$ metrics are even superior compared to the models from the test dataset. Both shallow thin resistive targets are restored well; the position of the deeper target is also accurate although its resistivity is underestimated by ~30-45 Ohm-m. The uncertainty panel on Figure 5 shows the maximum difference in the predictions of the networks from 10 different epochs (scaled by the average resistivity value in the corresponding cell; the average is taken over the predictions from 10 epochs and the true model).

*4.2. Time-domain inversion*

In this section, we examine the performance of the method on an ABEM WalkTEM dataset provided by Guideline Geo. The data was collected in Denmark and consists of 15 soundings along two survey lines. Each of the soundings consists of two data segments: low moment and high moment data. The transmitter coil is a 40x40 m loop and the ABEM RC5 receiver has an effective area of 5 m$^2$ and amplification of a factor of 7. The CNN is trained on the synthetic TEM dataset described in Section 3. The input data is much smaller compared to the previous case and includes the induced voltage $V$ measured at 34 time gates in a vertical induction coil. The convolutional part of the network has three levels (N is chosen as 8) and is followed by 5 dense layers of 128 neurons each and the output layer of 50 neurons.

In Figure 6, we compare our inversion results with the conventional inversion of the same WalkTEM dataset as 2D sections along Lines 1 and 2. The conventional Levenberg-Marquardt inversion determines 9 parameters for each sounding (5 resistivity values and 4 thicknesses of the layers), while the CNN outputs 50 resistivity values for much thinner layers. Figure 7 shows the same comparison for each individual sounding from the WalkTEM dataset and for several synthetic models from the synthetic test dataset. The average normalized RMSE achieved by the network on the latter test set is 0.121. For the real examples, the error reaches 0.196 (when compared to the results of deterministic inversion) which is expected since the "real" models come from another distribution (5 layered-models versus 50-layered models used in training); nevertheless, the quality of inversion is quite good as shown in Figures 6 and 7.

## 5. Discussion and conclusions

Traditional EM inversion methods are commonly based on gradient-based algorithms, which suffer from non-linearity and non-uniqueness of the inverse problem. On the other hand, the use of probabilistic inversion methods is often limited due to their high computational cost. A possible alternative lies in inversion based on deep neural networks, which are capable of learning deep representations and identifying complex patterns in data. This is of particular importance for large real datasets where complex data interactions are difficult or even impossible to specify within existing models. In this study, we explored the potential of DL based inversion with two 1D EM problems and showed that deep CNN can accurately reconstruct the resistivity distribution in the subsurface from measured data.

The workflow of our DL inversion consists of three stages: data generation, network training and the prediction itself. The first two stages are computationally expensive, since the networks must be trained on sufficiently large training sets in order to generalize well to new data. However, they are only performed once (offline). The inversion can be performed instantaneously (typically within in a few milliseconds per model) and, as demonstrated by the examples in this study, allows for sufficiently accurate estimation of the subsurface parameters. For the frequency-domain CSEM scenario, we observe that shallow resistive and conductive structures do not significantly impede the detection of deeper targets. The DL inversion of the real TEM data yields similar results to those of a conventional inversion. For multidimensional problems, DL inversion can potentially provide sufficiently accurate results orders of magnitude faster than conventional inversion methods. We also note that the training dataset can be easily extended further by adding random noise to the existing examples, thus making the method closer to real-world conditions. The performance of the CNN and smoothness of the resulting model can be tuned by using a different loss function instead of the RMSE.

First applications of EM inversion with deep neural networks suggest that it will supplement or replace traditional methods for inversion of model parameters from measured observations in the future. The results of this study suggest that a framework with various DL methods for 2D and 3D EM inversion is feasible.

**Acknowledgements**

The authors acknowledge support from the Curtin University Oil and Gas Innovation Centre (CUOGIC), the Curtin Institute for Computation (CIC), and the Institute for Geoscience Research (TIGeR). This work was supported by resources provided by the Pawsey Supercomputing Centre with funding from the Australian Government and the Government of Western Australia. VP greatly appreciates fruitful discussions with Shiv Meka on various aspects of artificial intelligence. The authors also thank Guideline Geo for providing the ABEM WalkTEM demonstration dataset from Denmark.

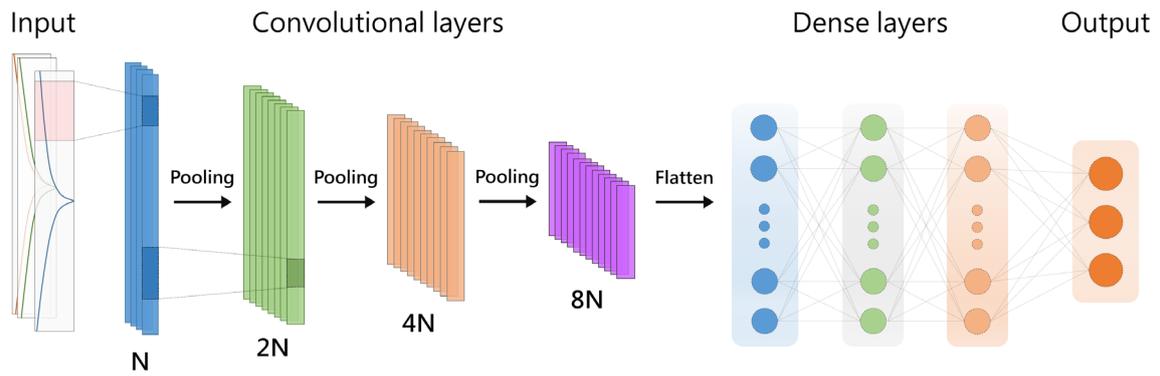

**Figure 1.** Architecture of the neural networks. Following VGG-like and ResNet-like 1D CNN design (Simonyan and Zisserman, 2014; He et al., 2016), each convolutional layer consists of a block of two or three convolutions with the same number of filters (N, 2N, 4N or 8N) followed by the batch normalization operation. Each convolution operation has a kernel size of 5 and uses the 'same' padding. The number of dense layers and the number of neurons in the dense and output layers is specified below for each particular case.

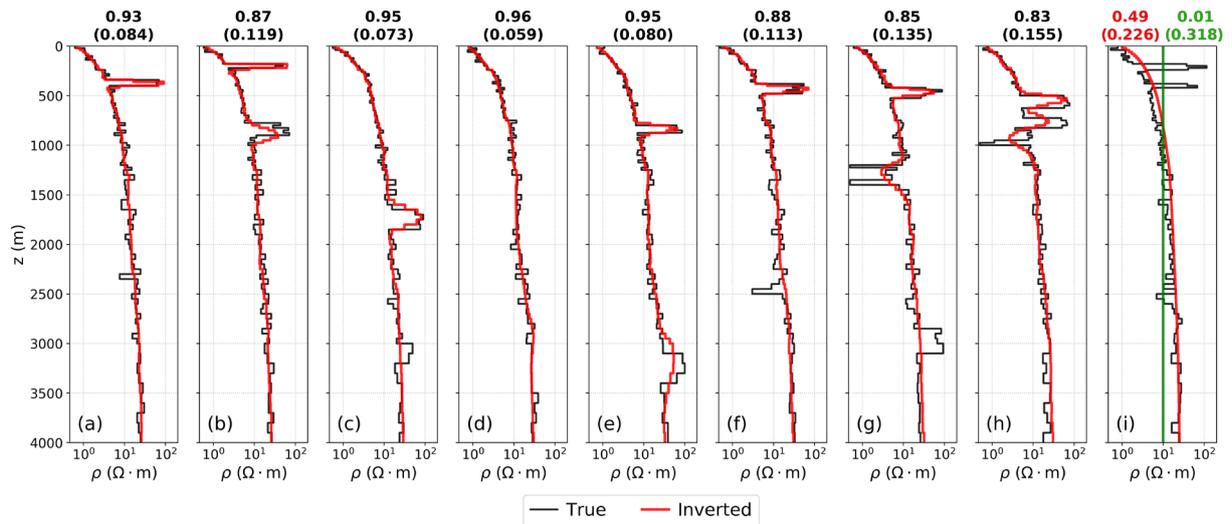

**Figure 2.** Results of the CNN inversion (red) versus the true resistivity models (black) for eight random examples from the test dataset. Five frequencies of 0.1, 0.3, 1, 3, and 10 Hz were used. The $R^2$ and normalized RMSE values for each example are shown in the title. The average $R^2$ and normalized RMSE for the entire test dataset are 0.912 and 0.093, respectively. Example (i) shows these metrics for a model with gradually increasing resistivity (red) and a 10 Ohm-m half-space.

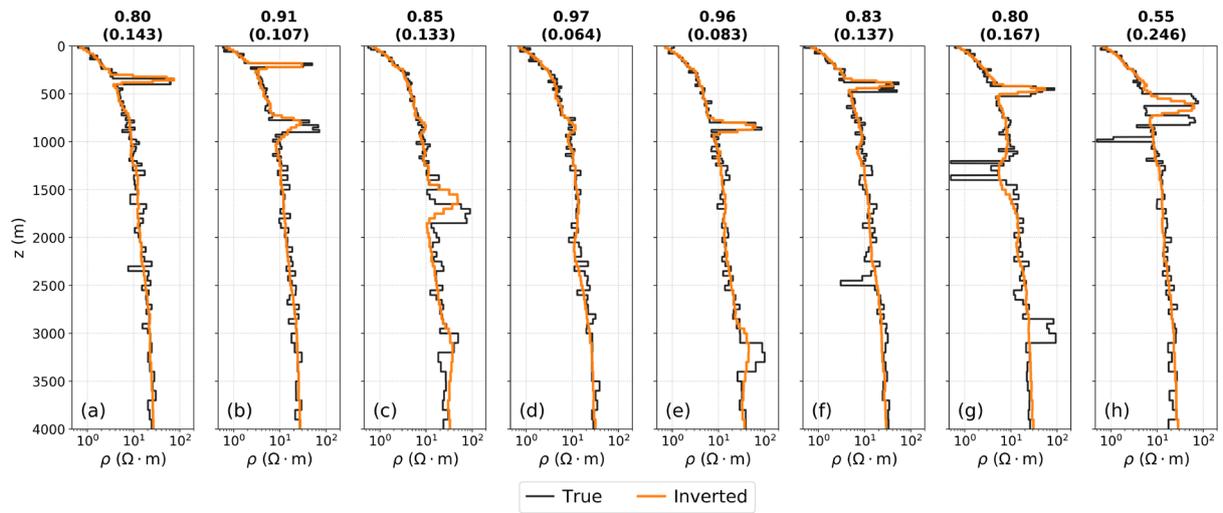

**Figure 3.** Results of the CNN inversion (orange) versus the true resistivity models (black) from the test dataset. Two frequencies of 0.1 and 1 Hz were used. The $R^2$ and normalized RMSE values for each example are shown in the title. The average $R^2$ and normalized RMSE for the entire test dataset are 0.863 and 0.121, respectively.

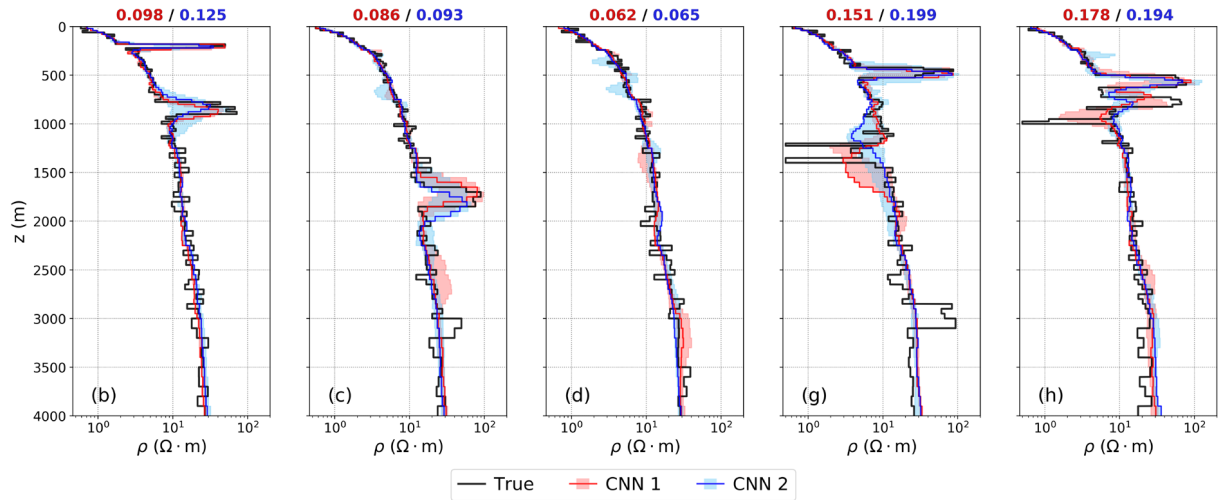

**Figure 4.** Inversion results using a CNN with N=80 and kernel size equal to 5 (red) and a simpler CNN with N=40 and kernel size equal to 3 (blue) versus the true resistivity models (black). All five CSEM frequencies were used, as in figure 2. The average normalized RMSE for the 10 epochs are shown in the titles for the first and the second networks, respectively. The solid lines are the predictions at the last epoch and the contours show the range.

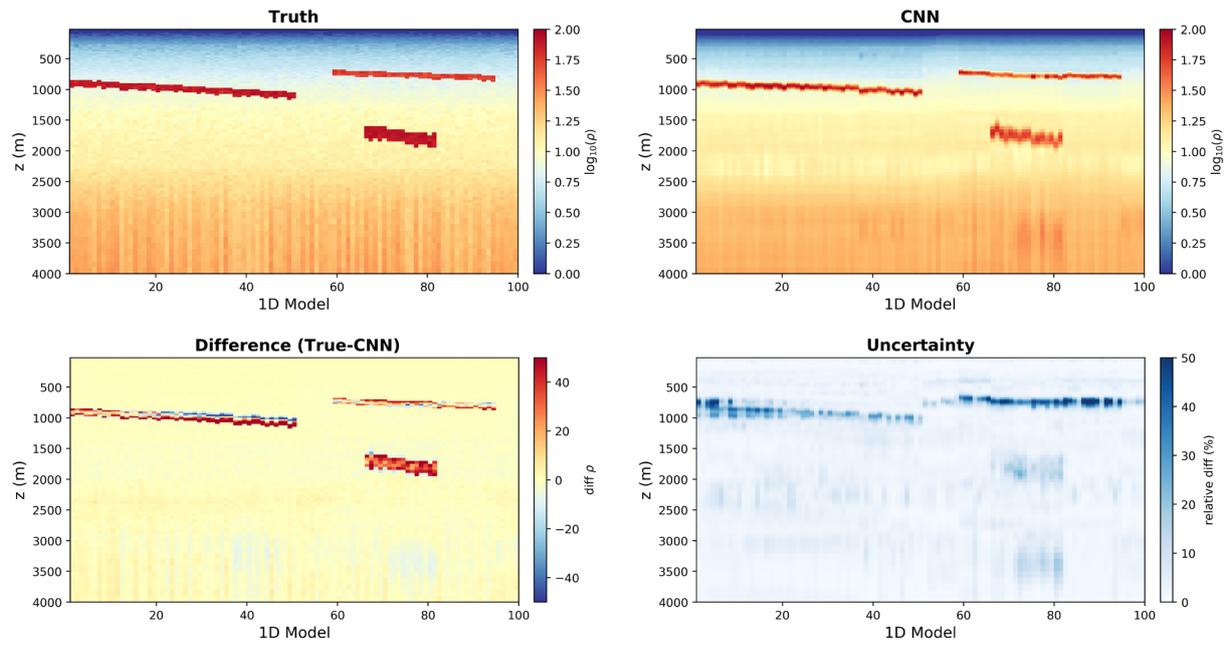

**Figure 5.** Top row: true 2D model comprised of 100 resistivity profiles (left) and results of the CNN inversion (right) on logarithmic scale. Bottom row: absolute difference between the true and inverted models on linear scale (left) and maximum relative difference in the predictions of the networks from 10 different epochs (right). Five frequencies of 0.1, 0.3, 1, 3, and 10 Hz were used. The average $R^2$ and normalized RMSE for these 100 examples are 0.93 and 0.085, respectively.

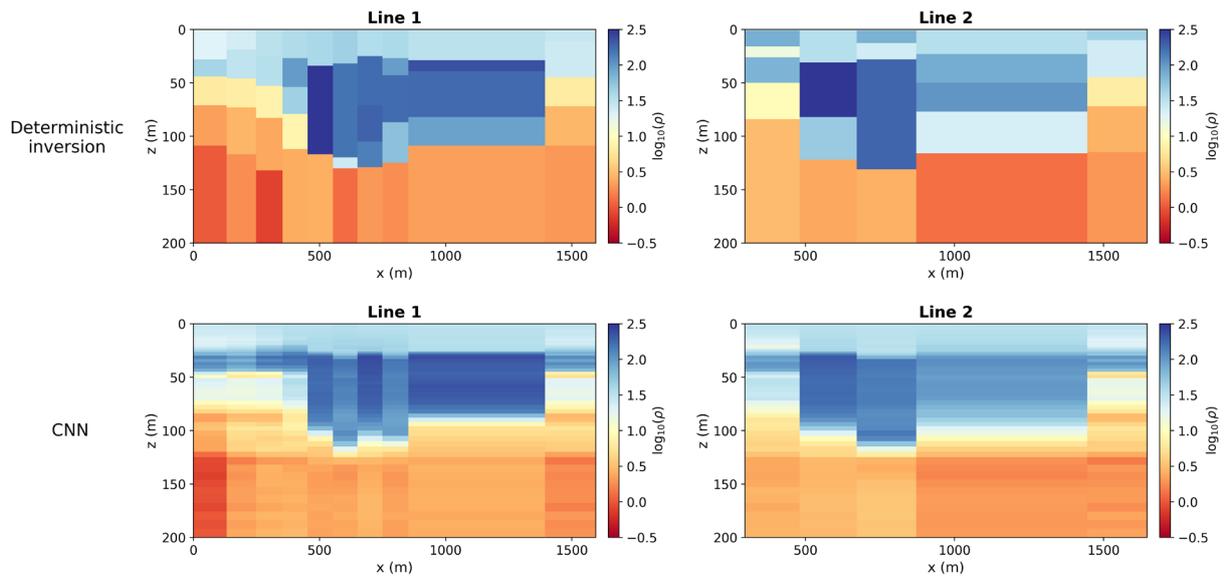

**Figure 6.** Top row: results of the conventional inversion for 5 layers with free boundaries for Line 1 (left) and Line 2 (right). Bottom row: results of the CNN inversion for 50 layers of fixed thickness for Line 1 (left) and Line 2 (right). Red color denotes high conductivity.

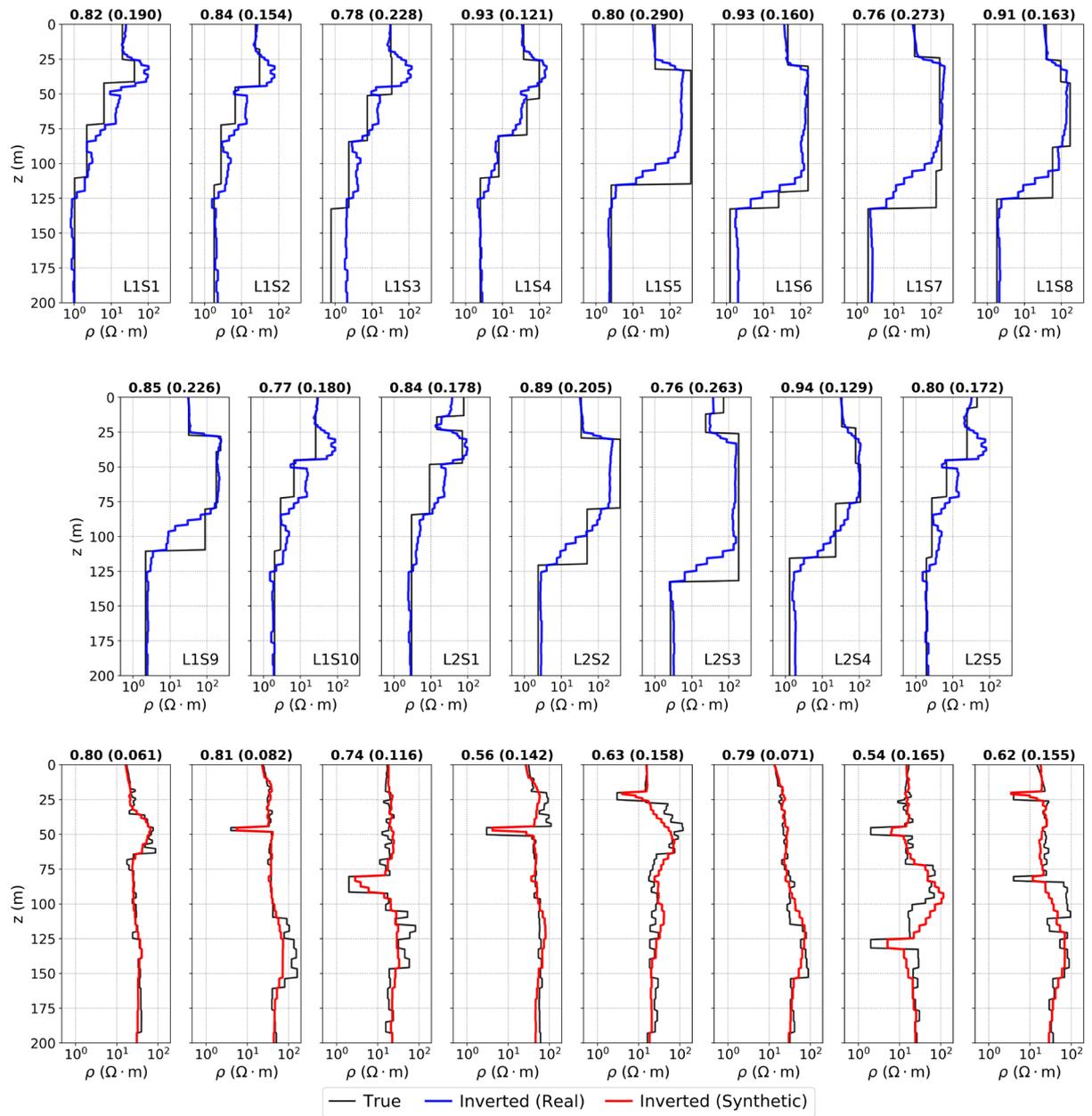

**Figure 7.** Top and middle rows: results of the CNN inversion (blue) versus the corresponding Levenberg-Marquardt inversion (black) for the real WalkTEM data. Bottom row: results of the CNN inversion (red) versus the corresponding true resistivity models (black) for synthetic TEM data. The $R^2$ and normalized RMSE values for each example are shown in the title.

| Anomaly | Probability of existence | Resistivity values (Ohm-m) | | Thickness range (m) | Location in the model |
|---|---|---|---|---|---|
| | | Mean | Deviation | | |
| **R1** | 0.9 | 50 | 20 | 20–60 | Top 1000 m |
| **R2** | 0.5 | 40 | 15 | 30–80 | Anywhere |
| **R3** | 0.1 | 30 | 12 | 20–100 | Anywhere |
| **R4** | 0.01 | 100 | 40 | 20–60 | Anywhere |
| **C1** | 0.2 | –6 | 4 | 50–250 | Top 2500 m |

**Table 1.** Parameters of the resistive (R) and conductive (C) anomalies used in the synthetic models. The negative value of the conductive anomaly arises by subtracting the resistivities of the corresponding layers.

| Training dataset size | Number of epochs | CNN error | | NN error | |
|---|---|---|---|---|---|
| | | Training | Test | Training | Test |
| 10,240 | 1000 | 0.071 | **0.129*** | 0.069 | 0.132* |
| 50,176 | 200 | 0.092 | **0.114** | 0.091 | 0.121 |
| 102,400 | 100 | 0.095 | **0.106** | 0.097 | 0.113 |
| 501,760 | 20 | 0.089 | **0.093** | 0.105 | 0.108 |

**Table 2.** RMSE errors for varying training dataset size. The test dataset always consists of 5120 examples.

* Overfitting is observed; lower test errors observed at previous epochs are around 0.125.